\documentclass[amsmath,amssymb,pra,aps,showpacs,superscriptaddress,twocolumn]{revtex4-2}
\usepackage{amsmath,amsfonts,amssymb,amsthm,graphics,graphicx,epsfig,bbm}
\usepackage[colorlinks=true,citecolor=blue,linkcolor=blue,urlcolor=blue]{hyperref}
\usepackage[usenames]{color}
\usepackage{graphicx}
\usepackage{subfigure}
\usepackage{amsmath}
\usepackage{epsfig}
\usepackage{dcolumn}
\usepackage{bm}
\usepackage{color}
\usepackage{times}
\usepackage{epstopdf}
\usepackage{amssymb}
\usepackage{amstext}
\usepackage{latexsym}
\usepackage{hyperref}
\usepackage{amsfonts}
\usepackage{psfrag}
\usepackage{soul,xcolor}
\usepackage[normalem]{ulem}
\usepackage{dsfont}
\usepackage{txfonts}
\usepackage{float}
\usepackage{makecell}
\usepackage{comment}
\usepackage{appendix}
\usepackage{diagbox}
\usepackage{tikz}

\usepackage{graphicx}
\usepackage{dcolumn}
\usepackage{bm}

\newcommand{\ket}[1]{\vert #1 \rangle}
\newcommand{\bra}[1]{\langle #1 \vert}

\newcommand{\braket}[2]{\langle #1 \vert #2 \rangle}

\newcommand{\etal}{{\it{et al.}}}

\begin{document}

\title{Quantum coherence and counterdiabatic quantum computing}

\author{Raziel Huerta-Ruiz}
\affiliation{Facultad de Ciencias, Universidad Nacional Autónoma de México (UNAM), Av. Universidad No. 3000, Col. Universidad Nacional Autónoma de México, C.U., Delegación Coyoacán, C.P. 04510, Ciudad de México, México}

\author{Maximiliano Araya-Gaete}
\affiliation{Departamento de Física, Universidad de Santiago de Chile (USACH), Avenida Víctor Jara 3493, 9170124 Santiago, Chile}

\author{Diego Tancara}
\affiliation{Departamento de Física, Universidad de Santiago de Chile (USACH), Avenida Víctor Jara 3493, 9170124 Santiago, Chile}

\author{Enrique Solano}
\affiliation{Kipu Quantum, Greifswalderstrasse 212, 10405 Berlin, Germany}

\author{Nancy Barraza}
\affiliation{Universidad Diego Portales, Cedenna, Ejército 441, Santiago, Chile}

\author{Francisco Albarr\'an-Arriagada }
\affiliation{Departamento de Física, Universidad de Santiago de Chile (USACH), Cedenna, Avenida Víctor Jara 3493, 9170124 Santiago, Chile}
\email{francisco.albarran@usach.cl}

\date{\today}

\begin{abstract}

Counterdiabatic driving emerges as a valuable technique for implementing shortcuts to adiabaticity protocols, enhancing quantum technology applications. In this context, counterdiabatic quantum computing represents a new paradigm with the potential to achieve quantum advantage for industrial problems. This work investigates the production of quantum coherence in adiabatic evolution accelerated by counterdiabatic driving within the framework of counterdiabatic quantum computing. Specifically, we analyze different orders in the nested commutator expansion for approximated counterdiabatic drivings for three cases: a weighted max-cut problem, a 4-local Hamiltonian, and a non-stoquastic Hamiltonian. Our findings reveal that the hierarchy introduced by coherence production correlates with the success probability in the impulse regime. This suggests that protocols increasing coherence during evolution enhance performance in adiabatic evolution driven by counterdiabatic techniques.  We show that large quantum coherence also means large energy fluctuation during evolution, which is associated with the speed of evolution, paving the way for designing superior algorithms in counterdiabatic quantum computing.
\end{abstract}

\maketitle


\section{Introduction}

Quantum computing has the potential to surpass the capabilities of classical devices, as demonstrated by several experiments claiming quantum advantage \cite{Arute2019Nature, Zhong2021PhysRevLett, Wu2021PhysRevLett, Madsen2022Nature, Kim2023Nature, Zhu2022SciBull, Morvan2024Nature, Acharya2024Nature, Gao2025PhysRevLett, King2025Science}. However, many of these experiments were designed for boson sampling~\cite{Aaronson2011ACM} and random circuit sampling~\cite{Boixo2018NatPhys, Bouland2019NatPhys}, two tasks whose potential use in industrial applications is not straightforward. Global efforts to achieve quantum advantage in industrial cases have proposed new paradigms to create more efficient quantum algorithms. One such paradigm is adiabatic quantum computing enhanced by shortcuts to adiabaticity techniques, known as counterdiabatic quantum computing (CQC).

As in adiabatic quantum computing, CQC perform the evolution from an initial Hamiltonian, whose ground state is easily prepared, to a final Hamiltonian, where the ground state represents the solution to the optimization problem. The main difference between CQC and adiabatic quantum computing is the addition of counterdiabatic driving (CD) terms. CD terms aim to suppress undesired transitions between instantaneous eigenstates, mimicking the adiabatic path in a shorter time~\cite{OdelinRevModPhys2019, DemirplakJPhysChemA2003, DemirplakJPhysChemA2005, Berry2009JPhysA}, theoretically accelerating an adiabatic process as much as desired. Despite this potential, the exact computation and implementation of counterdiabatic fields are generally challenging, as they require knowledge of the instantaneous eigenstates of the adiabatic Hamiltonian throughout the entire evolution, making it impractical in most cases. In response, approximate expressions for the CD terms have been proposed, offering a more feasible path for implementing CD protocols~\cite{SelsPNAS2017, HatomuraPhysRevA2021, ClaeysPhysRevLett2019, MorawetzPhysRevB2024, CepaitePRXQuantum2023, FinžgararXiv2025, MorawetzarXiv2025, ShinguarXiv2025}. In this context, CQC can be implemented in a digital form~\cite{HegadePhysRevResearch2022}, showing great potential for various applications such as factorization, biological systems, financial modeling, and more \cite{HegadePhysRevRes2022, HegadePhysRevA2021, ChandaranaPhysRevRes2022, ChandaranaPhysRevAppl2023, CadavidPhysRevAppl2024, GuanQuantumSciTechnol2024, ChandaranaCommunPhys2024}. Additionally, CQC has been implemented in the analog paradigm in cold atoms~\cite{ZhangarXiv2024}, demonstrating the flexibility of CQC to adapt to different available quantum devices. Even though CQC represents a novel approach to fast and robust optimization, the complete understanding of the quantum resources exploited by CQC is not entirely known. 

Among the various quantum resources, quantum coherence is particularly interesting because it quantifies the superposition in a given basis, generally defined by Hamiltonian of the system. CD terms modifies the instantaneous Hamiltonian to follow an adiabatic path quickly, changing the coherence in the instantaneous energy basis. Therefore, it is natural to consider that coherence production during evolution plays a fundamental role. The role of quantum coherence has been studied in different contexts such as quantum speed limit~\cite{MarvianPhysRevA2016, XuPhysRevRes2020, MaiPhysRevA2024}, quantum synchronization~\cite{SolankiPhysRevA2022}, quantum thermodynamics~\cite{GourPRXQuantum2022}, quantum reservoir computing~\cite{PalaciosCommunPhys2024}, and quantum algorithms~\cite{HilleryPhysRevA2016, AhnefeldPhysRevLett2022, LiPhysLettA2018}, but to the best of our knowledge, it has never been studied within the framework of CQC.

In this work, we study the role of quantum coherence in CQC. Specifically, we consider the production of quantum coherence and its relation to the accuracy in ground state preparation for different problems solved by CQC: the weighted max-cut problem, a 4-local Hamiltonian, and a non-stoquastic Hamiltonian, with non-trivial entanglement in its ground state, given by the Heisenberg model. We also relate this production to energy fluctuation during evolution, which is strictly related to the speed of evolution, providing a perfect hierarchy for approximations to counterdiabatic driving, in our case, the nested commutator expansion given by Ref.~\cite{ClaeysPhysRevLett2019}. With this work, we aim to establish a framework to understand the role of quantum coherence in counterdiabatic driving for quantum computing.

\section{Methods}

\begin{figure}[b]
\centering
    \includegraphics[width=0.4\linewidth]{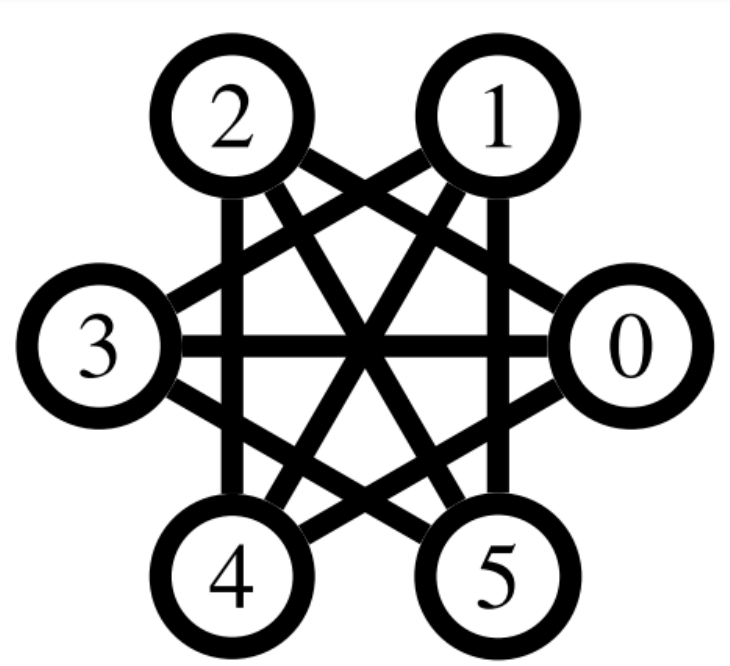}    
    \caption{Graph $\mathcal{G}$ for max-cut problem example.}\label{Fig01}
\end{figure}

This section briefly describes the basics of adiabatic evolution and counterdiabatic driving, as well as coherent measurements and energy fluctuations.

\subsection{Adiabatic quantum computing and counterdiabatic drivings}

Adiabatic quantum computing emerges as a method to access the ground state of a Hamiltonian from another by an adiabatic evolution governed by the following time-dependent Hamiltonian: 
\begin{equation} 
H_{\text{ad}} (t) = H_{I} (1-\lambda(t)) + H_{F} \lambda(t), 
\label{Eq01} 
\end{equation} 
where $\lambda(t)$ is the schedule function with $\lambda(0) = 0$, $\lambda(T) = 1$, and $T$ is the final time of the evolution~\cite{AlbashRevModPhys2018}. Although the adiabatic theorem guarantees that there is a time $T$ such that the final state of the evolution will be close enough to the ground state of $H_F$ if the initial state of the evolution is the ground state of $H_I$, this time could be longer than the coherence time of the quantum system, making the adiabatic evolution impractical. To address this, CD driving accelerates convergence to the final ground state by introducing an additional term to the adiabatic Hamiltonian. In general, we can write the counterdiabatic Hamiltonian as:
\begin{equation} 
H(t) = H_{\text{ad}}(t) + \dot{\lambda} A_{\lambda}(t), 
\label{Eq02} 
\end{equation} 
where $A_{\lambda}(t)$ is known as the adiabatic gauge potential (AGP) and takes the form: 
\begin{equation} 
A_{\lambda}(t) = -i\hbar \sum_{n\ne m} \frac{\langle m(t)|\partial_{\lambda}H_{\text{ad}}(t)|n(t)\rangle}{\epsilon_m(t)-\epsilon_n(t)} |m(t)\rangle \langle n(t)|, 
\label{Eq03} 
\end{equation} 
where $|n(t)\rangle$ and $\epsilon_n(t)$ are the $n$th instantaneous eigenstates and eigenenergy of $H_{\textrm{ad}}(t)$, respectively, such that $ H_{\textrm{ad}} (t) |n (t)\rangle = \epsilon_n(t) |n (t)\rangle $. As Eq. (\ref{Eq03}) requires knowledge of the spectrum of $H_{\textrm{ad}}(t)$ at each time, making the exact calculation of $A_\lambda(t)$ impractical. Nevertheless, we can calculate approximate versions of the AGP without the spectral information of $H_{\textrm{ad}}(t)$. One of such approximation is given in Ref.~\cite{ClaeysPhysRevLett2019} and is based in the next series of nested commutators: 
\begin{equation} 
A_{\lambda} (t) = i \sum_{k = 1}^{\infty} \alpha_k(t) O_{2k-1} (t), \label{Eq04} \end{equation} \begin{equation} O_k (t) = \underbrace{[H_{\text{ad}},[H_{\text{ad}}, \cdots, [H_{\text{ad}}}{k}, \partial\lambda H_{\text{ad}}]]], 
\label{Eq05} 
\end{equation} 
where $\alpha_k$ can be determined by solving the following linear system of equations \cite{XiePhysRevB2022}: \begin{equation} 
\begin{bmatrix} 
\Gamma_2 & \Gamma_3 & \cdots & \Gamma_{l+1} \\
\Gamma_3 & \Gamma_4 & \cdots & \Gamma_{l+2} \\
\vdots & \vdots & \ddots & \vdots \\
\Gamma_{l+1} & \Gamma_{l+2} & \cdots & \Gamma_{2l} 
\end{bmatrix} 
\begin{bmatrix} \alpha_1 \\ 
\alpha_2 \\ 
\vdots \\ 
\alpha_l 
\end{bmatrix} 
= - \begin{bmatrix} \Gamma_1 \\ \Gamma_2 \\ \vdots \\ \Gamma_l 
\end{bmatrix}, 
\label{Eq06} 
\end{equation} 
with $\Gamma_k = ||O_k ||^2_\text{F} $ being the square of the Frobenius norm of the nested commutator, defined as $|| A ||_\text{F} = \sqrt{\text{Tr}(A^\dagger A)}$. If we truncate this infinite series, we can define the $l$th order approximation by: 
\begin{equation} 
A_{\lambda}^{(l)} (t) = i \sum_{k = 1}^{l} \alpha_k(t) O_{2k-1} (t). 
\label{Eq07} 
\end{equation}

We note that this approximation, as well as the exact form of the adiabatic gauge potential, depends on the change of the Hamiltonian with the scheduling functions, and its strength is proportional to the change of the scheduling function in time ($\dot{\lambda}$), implying that the counterdiabatic driving will be determinant for short evolution time $T$ (fast changes), known as the impulse regime~\cite{CarolanPhysRevA2022}, and less important for long evolution time (slow changes), called the adiabatic regime. In this sense, we can consider fast changes when the change of the scheduling function is much larger than the energy gap between the ground state and the first excited state ($\Delta$), which is proportional to the inverse of the minimal time scale of the system. Then, the impulse regime can be defined by: 
\begin{equation} 
\max_{t \in [0,T]} \left ( \frac{\dot{\lambda}(t)}{\Delta} \right ) \gg 1. 
\label{Eq08} 
\end{equation}
If we consider: 
\begin{equation} 
\lambda(t) = \sin^2 \left ( \frac{\pi}{2} \sin^2 \left (\frac{\pi t}{2 T} \right ) \right ), \label{Eq09} \end{equation} then: \begin{equation} \dot{\lambda}(t) = \frac{\pi^2}{4 T} \sin \left ( \frac{\pi t}{T} \right ) \sin \left ( \pi \sin^2 \left (\frac{\pi t}{2 T} \right ) \right ), 
\label{Eq10} 
\end{equation} 
obtaining the following condition for the final time $T$ in the impulse regime: 
\begin{equation} 
\frac{\pi^2}{4} \gg T\Delta. \label{Eq11} 
\end{equation}
Using a similar analysis, we can define the condition for the adiabatic regime as: 
\begin{equation} 
\frac{\pi^2}{4} \ll T \Delta, 
\label{Eq12} 
\end{equation} 
and an intermediate regime when $\frac{\pi^2}{4} \sim T\Delta$.

\subsection{Quantum Coherence and Energy Fluctuation}

In this work, we are interested in quantum superposition in the instantaneous eigenbasis generated by the adiabatic evolution boosted by CD driving. Quantum coherence quantifies such superposition. There are several measures of quantum coherence, generally based on the quantification of off-diagonal elements of the density matrix of a quantum state in a given basis~\cite{BaumgratzPhysRevLett2014}. Among the different measures of quantum coherence, we can mention trace distance~\cite{RanaPhysRevA2016}, robustness of coherence~\cite{NapoliPhysRevLett2016}, asymmetry~\cite{MarvianPhysRevA2016}, and relative entropy~\cite{BischofPhysRevLett2019}. In this work, we will consider the relative entropy of coherence defined as: 
\begin{eqnarray} 
C_{re}(t) = S(\rho_{diag}(t))-S(\rho(t)), 
\label{Eq13} 
\end{eqnarray} 
where $S(\cdot)$ refers to the von Neumann entropy, $\rho$ is the density matrix of our system, and $\rho_{diag}(t)$ is the diagonal part of $\rho(t)$. As we are interested in the superposition generated by the eigenbases of the system during the evolution, we consider the instantaneous energy bases, defined by Eq.~(\ref{Eq02}, to measure the quantum coherence. In our case, we will not consider dissipative effects during the evolution. Our quantum system is described by a pure state that in the instantaneous eigenbasis ($H(t)\ket{\phi_j(t)}=\alpha(t)\ket{\phi_j(t)}$) reads as: 
\begin{eqnarray} 
\ket{\Phi(t)}=\sum_j c_j(t)\ket{\phi_j(t)}, 
\label{Eq14} 
\end{eqnarray} 
implying that $S(\rho)=0$, and therefore: 
\begin{eqnarray} 
C_{re}(t) = S(\rho_{diag}(t))=-\sum_j |c_j(t)|^2\log_2\left(|c_j(t)|^2\right), 
\label{Eq15} 
\end{eqnarray} 
which is the Shannon entropy of the probability distribution of the instantaneous eigenstates of our system. As $C_{re}(t)$ is local in time, we can consider the mean coherence during the evolution defined by: 
\begin{equation}
C_P(T) = \frac{1}{T}\int_0^T C_{re}(t)dt. 
\label{Eq16} 
\end{equation}

An interesting feature of quantum coherence is that the corresponding energies of the eigenstates are not important. For this reason, it is also interesting to compute the energy fluctuation during the evolution, where the distribution of the eigenstates and the value of the eigenenergies play a main role. For the entire evolution, we will consider the average of the energy fluctuations given by: 
\begin{equation} 
\Delta\bar{E}(T) = \frac{1}{T}\int_0^T \sqrt{\bra{\Phi(t)} H^2(t)\ket{\Phi(t)}-\bra{\Phi(t)} H(t)\ket{\Phi(t)}^2}dt, 
\label{Eq17} 
\end{equation} 
which is related to the quantum speed limit as~\cite{Huang2023PhysRevRes}: 
\begin{equation} 
\tau_{QSL}=\frac{\arccos(\left| \braket{\Phi(0)}{\Phi(T)}\right|)}{\Delta \bar{E}(T)}. 
\label{Eq18} 
\end{equation}

\section{Results}

\begin{figure}[t]
    \centering
    \includegraphics[width=0.9\linewidth]{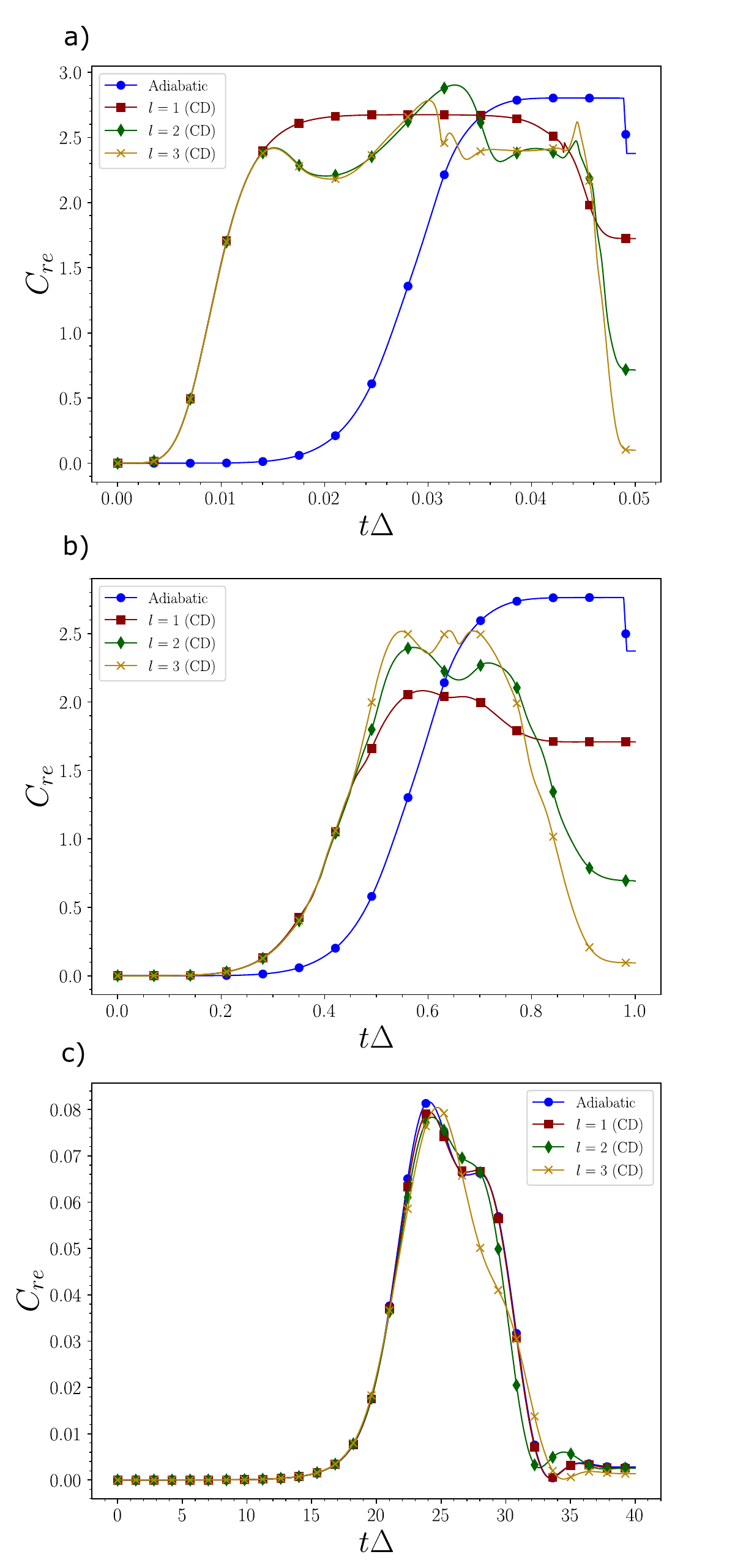}
    \caption{ Coherence $C_{re}$ as function of normalized time $\Delta_{gap} t$ for the max-cut problem defined by Fig.~\ref{Fig01}. a) Impulse regime  ($T\Delta << 1$). b) Intermediate regime ($T\Delta \sim 1$). c) Adiabatic regime ($T\Delta >> 1$).}
    \label{Fig02} 
\end{figure}

We will consider three kinds of problems. First, a quadratic unconstrained binary optimization (QUBO), which can codify the well-known max-cut problem. Second, we explore higher-order unconstrained binary optimization (HUBO), specifically a 4-local Hamiltonian that can codify a factorization problem. Finally, a non-stoquastic problem, specifically the Heisenberg model, in which the ground state has non-trivial entanglement.

\begin{figure}[b]
    \centering
    \includegraphics[width=1.0\linewidth]{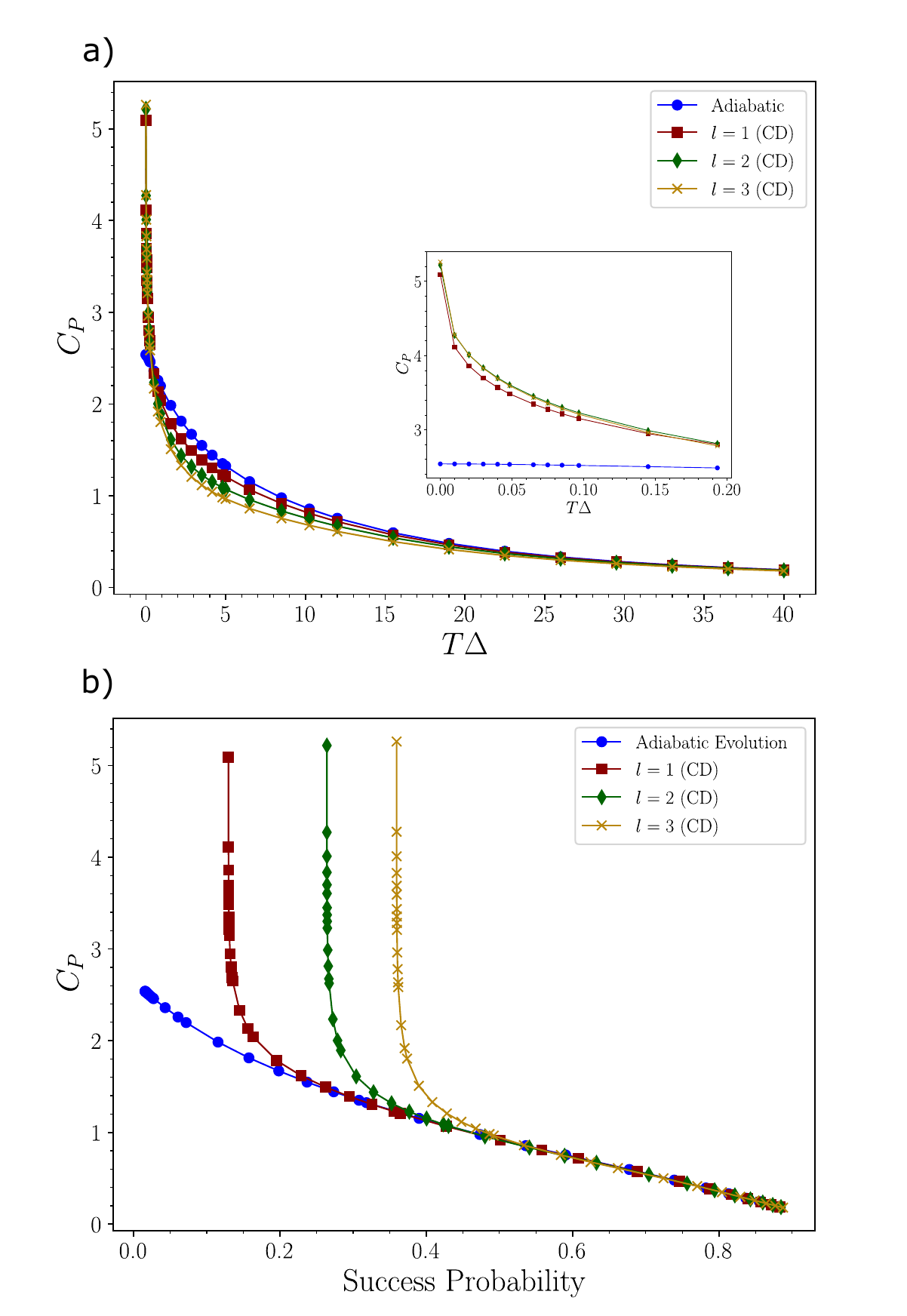}
    \caption{a) Mean coherence $C_P$ as function of total evolution time $T\Delta$, inset figure: impulse regime. b) Mean coherence $C_P$ as a function of success probability. Average results over 100 random QUBO Hamiltonians.}
    \label{Fig03} 
\end{figure}

\subsection{Quadratic Unconstrained Binary Optimization problems}
To develop some intuition, we first consider the maximum cut problem for the graph $\mathcal{G}=(V, E)$ composed of a group of vertices $V$ and edges $E$ given by Fig.~\ref{Fig01}. The graph has six vertices and nine edges, where each vertex is linked to the other three. In this case, the final problem Hamiltonian reads:
\begin{equation}
    H = \frac{1}{2}\sum_{\langle i,j\rangle\in E} J_{ij}(\sigma_i^z\sigma_j^z-1).
    \label{Eq19}
\end{equation}

Figure~\ref{Fig02} shows the quantum coherence in time for short evolution or impulse regime (a), intermediate regime (b), and long evolution time or adiabatic regime (c). For each regime, we consider the evolution under pure adiabatic Hamiltonian given by Eq. (\ref{Eq01}), and using counterdiabatic terms, considering the first three orders in the nested commutator expansion of the adiabatic gauge potential, see Eq. (\ref{Eq02}) and Eq. (\ref{Eq07}). We can observe that in the impulse regime, the counterdiabatic driving generates a rapid growth of the quantum coherence, which tends to disappear when the total evolution time goes to the adiabatic regime, as shown in Fig. \ref{Fig02}c. This suggests that a good metric for the total time evolution is the mean coherence given by Eq.~(\ref{Eq16}), the normalized area under the curve, which can introduce a hierarchy for counterdiabatic terms.

To test the apparent hierarchy provided by the quantum coherence, we can consider random instances of QUBO Hamiltonian, that is, Hamiltonians of the form: 
\begin{eqnarray} 
H = \sum_j \alpha_j \sigma_j^z + \sum_{j,k} \beta_{j,k}\sigma_j^z\sigma_k^z, 
\end{eqnarray} 
where the coefficients $\alpha_j, \beta_{j,k}$ are randomized in the interval ${-1, 1}$. We must mention that Eq.~(\ref{Eq19}) is a special case of a QUBO Hamiltonian.

Figure~\ref{Fig03} shows the average mean coherence for 100 random instances of a QUBO Hamiltonian. From panel a, we can observe in the inset figure that high-order CD terms are characterized by sizeable mean coherence in the impulse regime, which is inverted during the intermediate regime. In panel b, we can see that the success probability remains almost constant in the impulse regime, until the quantum coherence approaches the adiabatic regime. This suggests that the usefulness of approximated CD terms by nested commutator expansion appears only during the impulse regime and can be hierarchized by the mean coherence during the time evolution. 

\begin{figure}[t]
    \centering
    \includegraphics[width=1\linewidth]{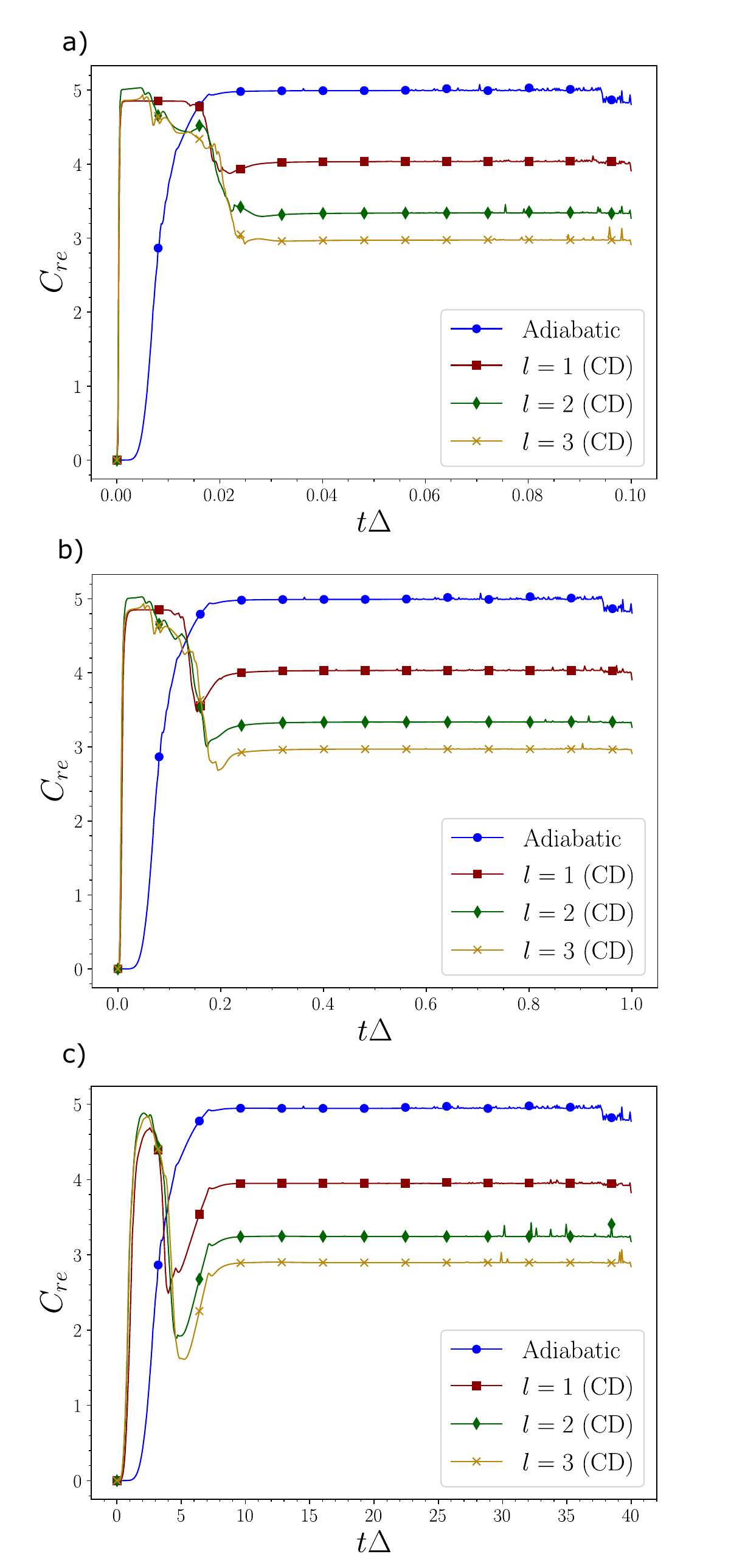}
    \caption{ Coherence $C_{re}$ calculated with relative entropy vs $\Delta t$ for the factorization problem of number 143. a) Impulse regime  ($\Delta T << 1$). b) Intermediate regime ($\Delta T = 1$). c) Adiabatic regime ($\Delta T >> 1$). }
    \label{Fig04} 
\end{figure}

\subsection{Higher-Order Unconstrained Binary Optimization}

Beyond quadratic optimization, we must address HUBO problems to solve any classical optimization. One pedagogical example is prime factorization, that is, to find two prime numbers, $x$ and $y$, whose multiplication equals $N$. This problem is equivalent to finding the minimal value of: 
\begin{equation} 
f(x,y)=(x \cdot y - N)^2. 
\end{equation}

Writing $x$ and $y$ in a binary representation, we get the HUBO representation of the prime factorization problem, where its corresponding Hamiltonian reads: 
\begin{eqnarray} 
H = \left( \left[\sum_{j=1}^{n_x}2^{j}Q_j+I \right] \left[\sum_{k=n_x+1}^{n_x+n_y}2^{k-n_x}Q_k+I \right]-N I\right)^2, 
\end{eqnarray} 
$n_x$ and $n_y$ are the qubits necessary for the binary representation of variables $x$ and $y$, respectively, which can be calculated as in Ref.\cite{HegadePhysRevA2021,PengPhysRevLett2008}. The operator $Q_j=(I-\sigma^z_j)/2$ is diagonal with eigenvalues $1$ and $0$. Specifically, we consider the example $N=143=11\cdot 13$. Figure~\ref{Fig04} shows the coherence production for this problem for the impulse, intermediate, and adiabatic regimes, where we can observe again that the coherence production for a short time is larger for high-order levels of the counterdiabatic driving. Still, later, the low orders, specifically the adiabatic case, generate more coherence.

\begin{figure}[b]
    \centering
    \includegraphics[width=0.9\linewidth]{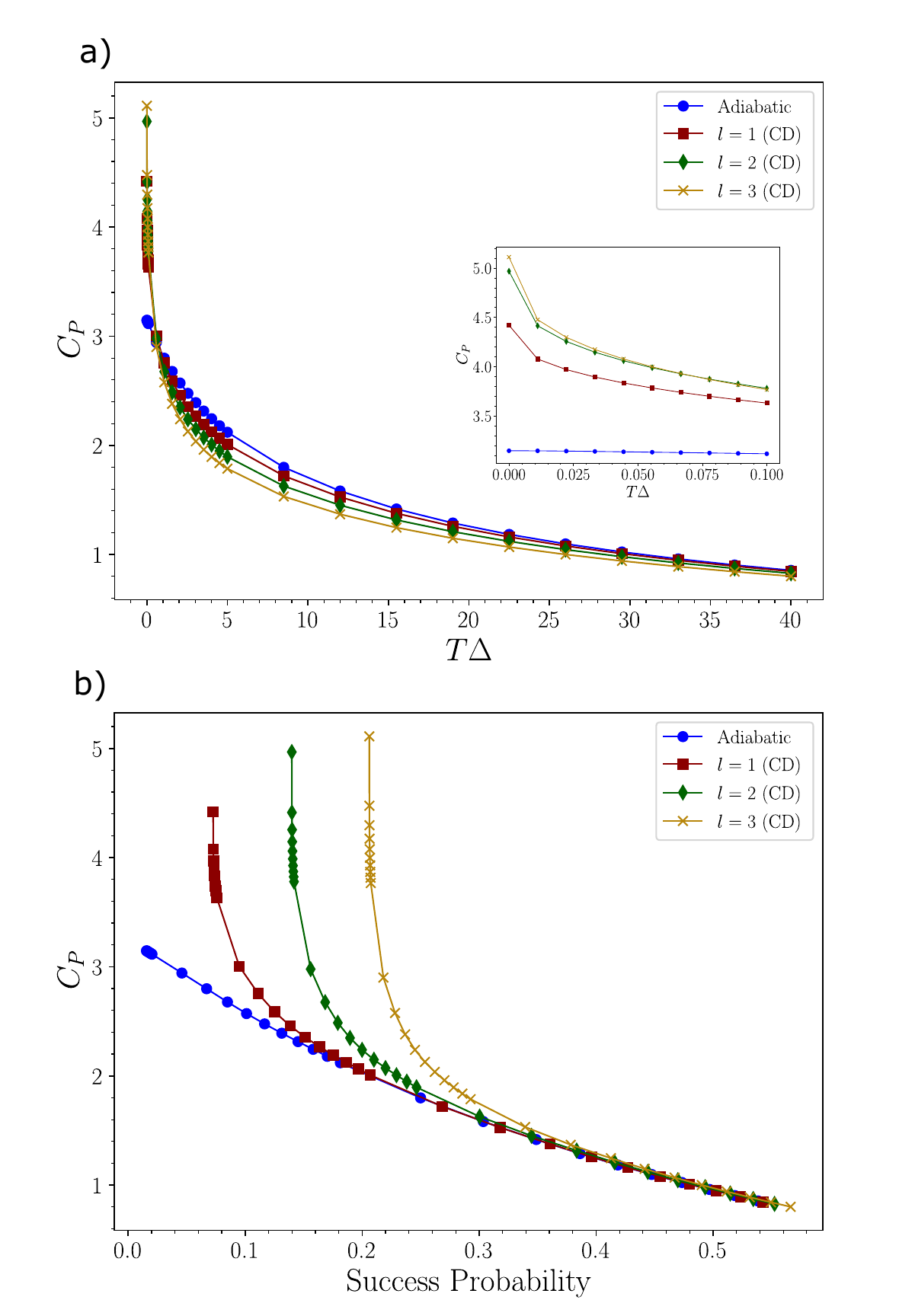}
    \caption{ 4-body Hamiltonian model for 100 random cases. a) Mean coherence $C_P$ vs time. b) Mean coherence vs success probability.}
    \label{Fig05} 
\end{figure}

Beyond the specific example of factorization, we can consider random instances for 4-local Hamiltonian of the form:
\begin{eqnarray}
    \notag H &=& \sum_j \alpha_j \sigma_j^z + \sum_{j,k} \beta_{j,k}\sigma_j^z\sigma_k^z + \sum_{j,k,l}\gamma_{j,k,l}\sigma_j^z\sigma_k^z\sigma_l^z\\  &&+ \sum_{j,k,l,m}\delta_{j,k,l,m}\sigma_j^z\sigma_k^z\sigma_l^z\sigma_m^z,
    \label{Eq23}
\end{eqnarray}
where the coefficients $\alpha,\beta,\gamma,\delta$ are randomized within the interval ${-1,1}$. Our previous example is a specific instance of Hamiltonian (\ref{Eq23}). Figure~\ref{Fig05} considers the mean coherence during the evolution for 100 random instances for different total evolution times and different final success probabilities. In this figure, we observe the change in the hierarchy of the adiabatic case concerning the evolutions with counterdiabatic driving. Also, as in the QUBO case, the success probability remains almost constant in the impulse regime, converging to the adiabatic case even if we consider high-order terms in the nested-commutator expansion.

Although the QUBO and HUBO Hamiltonians are very useful for coding problems with industrial applications, their ground states generally do not present quantum features. Therefore, it is interesting to consider a non-stoquastic final Hamiltonian whose ground state presents non-trivial quantum correlations.

\begin{figure}[t]
    \centering
    \includegraphics[width=0.9\linewidth]{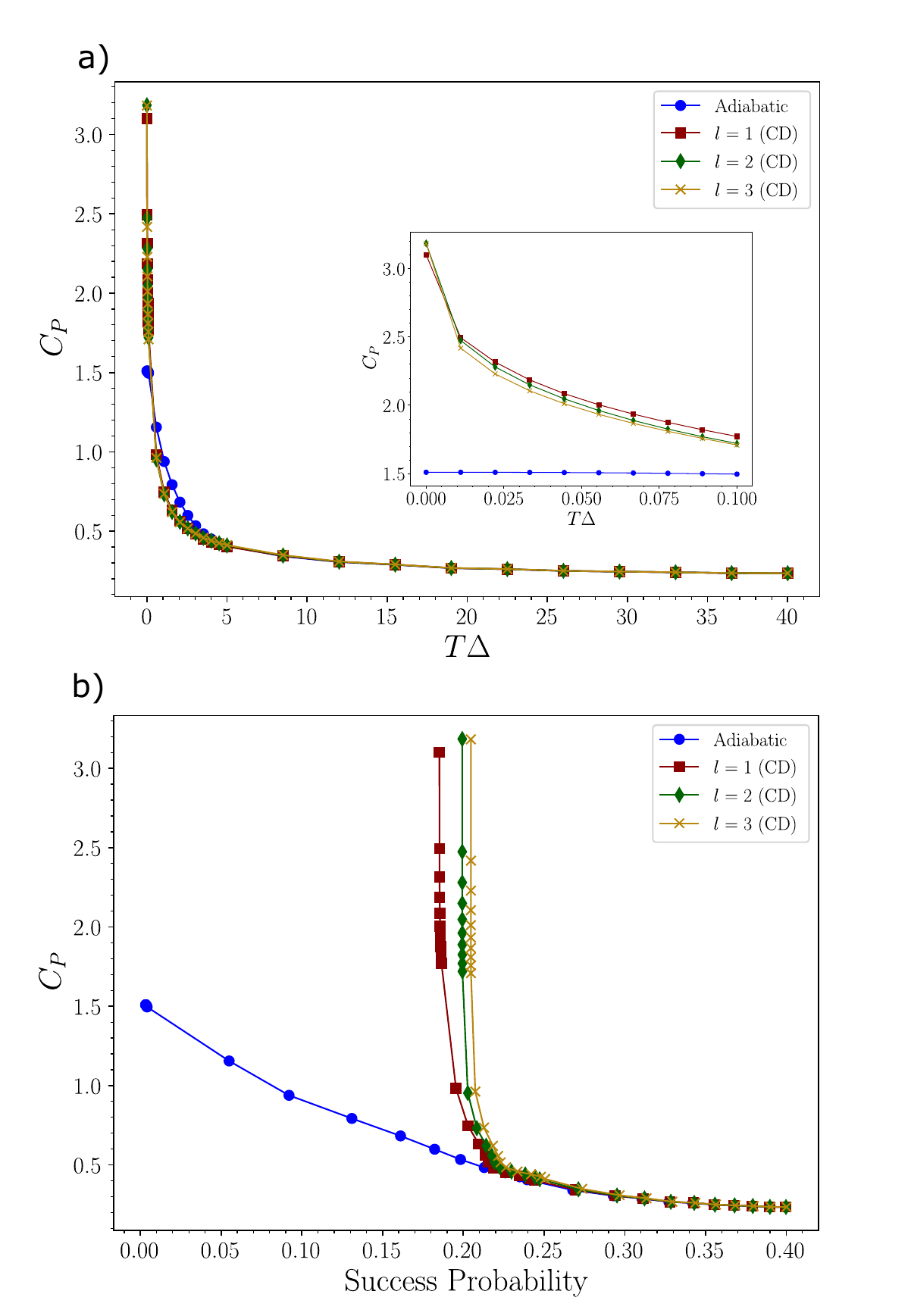}
    \caption{ Heisenberg model for 20 cases of equation \ref{eqHeisenberg}. a) Mean coherence $C_P$ vs time. b) Mean coherence vs success probability.}
    \label{Fig06} 
\end{figure}

\subsection{Heisenberg model}
We consider the Heisenberg model, which reads
\begin{equation}
    H = g\sum_{j=1}^N\sigma_j^z+J\sum_{j=1}^N\left(\sigma_j^x\sigma_{j+1}^x+\sigma_j^y\sigma_{j+1}^y\right)+\beta\sum_{j=1}^N\sigma_{j}^z\sigma_{j+1}^z,\label{eqHeisenberg}
\end{equation}
where $\sigma^{\alpha}_{N+1} = \sigma^{\alpha}_1$, that is, periodic boundary conditions. We consider $J/g=0.2$ and $\beta/g$ in the range $[0.2,0.8]$. This leads to entangled ground states, which represent a fundamentally different scenario from QUBO or HUBO problems, whose ground states generally do not exhibit entanglement or quantum correlations. From Fig.~\ref{Fig06}, we note again that the quantum coherence for a very short time (see inset in subfigure a), high orders of nested commutator tends to generate more quantum coherence which quickly changes, obtaining first that the evolution without CD terms produces less coherence compared to the evolutions with CD terms, which for the three first orders is almost equal. Later, as in the previous cases, the evolutions with CD terms start to produce less coherence, being the evolution without CD, the case that produces more quantum coherence for a long time.

Figure~\ref{Fig06}b shows how the quantum coherence is produced for a given success probability. As in previous cases, the impulse regime is characterized by an abrupt descent in the coherence, remaining almost constant the success probability until it reaches the intermediate regime and later the adiabatic one, where all the cases have practically the same behavior.

\section{Analysis and conclusions}
\begin{figure}[b]
    \centering
    \includegraphics[width=0.9\linewidth]{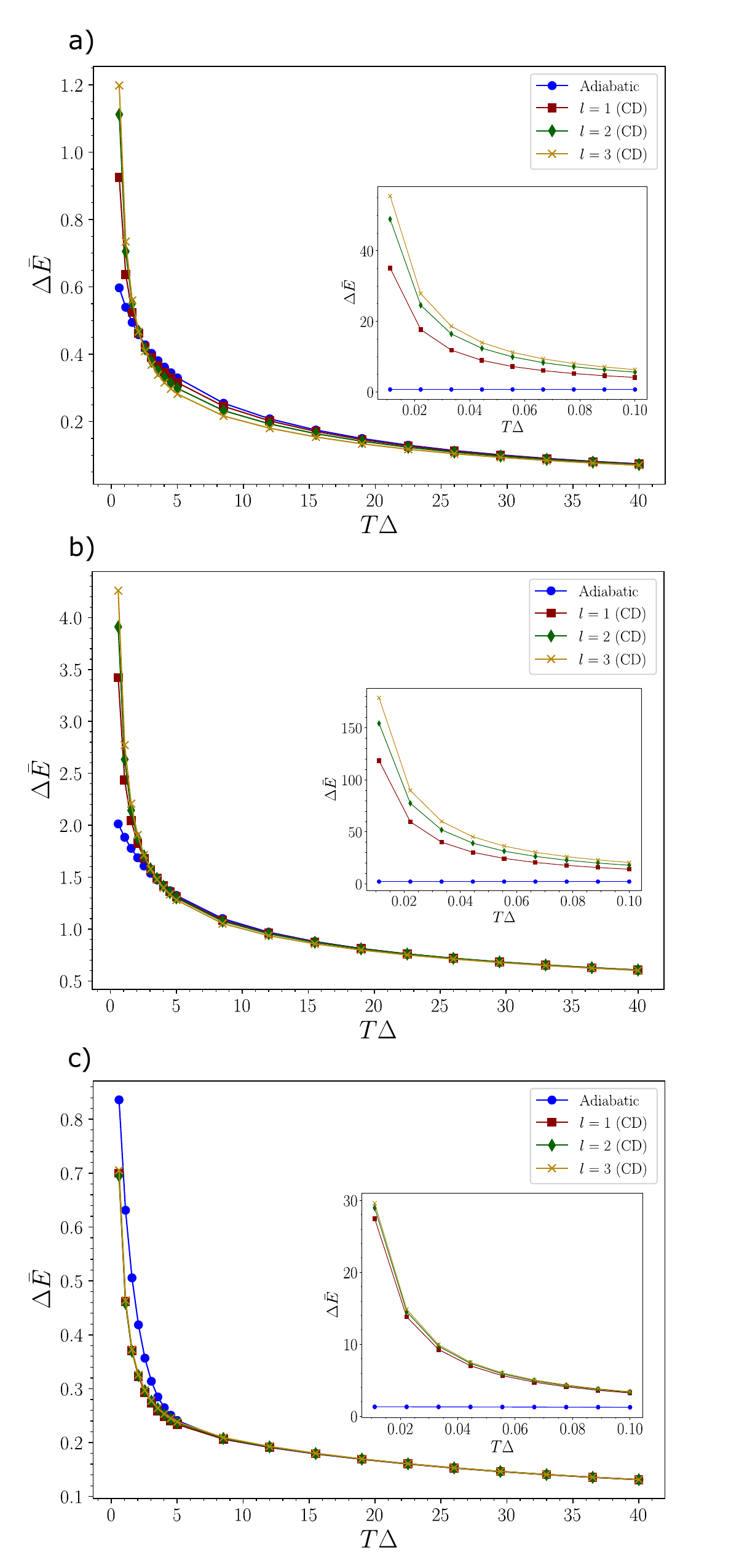}
    \caption{The average energy fluctuations as a function of the final time. We analyze the same set of random instances used for the mean coherence in a) the QUBO Hamiltonian, b) the 4-body Hamiltonian, and c) the Heisenberg Hamiltonian. The inset highlights the points within the impulse regime. }
    \label{Fig07} 
\end{figure}
The behavior of quantum coherence measured by $C_P$ shows that for all the examples considered for short times (impulse regime), complex approximations of the AGP produce more coherence. This means that in the impulse regime, better performance is associated with the capability to produce superposition between the different eigenstates of the Hamiltonian. This can be explained since in the impulse regime, the terms associated with the AGP are dominant in the total Hamiltonian, as they are proportional to the time derivative of the schedule function. Therefore, the only way to mimic the instantaneous ground state of the adiabatic Hamiltonian is by the superposition of eigenstates during the evolution, which is larger when we add more complex terms to the Hamiltonian, as reflected by the hierarchy shown in the different results. Additionally, we note that in the impulse regime, AGP approximations that produce more coherence, i.e., large order in Eq.~(\ref{Eq07}), have better performance according to the success probability, reinforcing the idea that large coherence production provides better performance in short evolution time.

On the other hand, for a long evolution time, i.e., the adiabatic regime, all evolutions provide approximately the same coherence production and success probability. This can be understood since in this regime, the time derivative of the schedule function is small, and therefore, the AGP can be considered as a perturbation, slightly changing the different measures related to the evolution. This means that for a long evolution time, the consideration of CD terms does not provide an appreciable enhancement in the performance of the evolution and the coherence tends to zero as in the adiabatic case.

Also, the behavior of coherence production can be related to the behavior of energy fluctuations, as shown in Fig. \ref{Fig07}, wherein the impulse regime, the energy fluctuations follow the same hierarchy of coherence production. That is, complex AGP provides more energy fluctuations, providing the mechanism to reduce the quantum speed limit for this class of evolutions, as can be observed from Eq. (\ref{Eq18}), opening the door to fast and accurate algorithms.

In conclusion, we have shown the relation and hierarchy between coherence production and the different approximations of the AGP, where better approximations produce more coherence and result in better performance. Additionally, this large coherence production is related to larger energy fluctuations, providing the physical mechanism to shorten the limits given by the quantum speed limit. We also mention that even if counterdiabatic evolution cannot reach a success probability larger than 0.5 in the impulse regime, these techniques can achieve enough success probability to recover the optimal solution by a sampling process with current technology, paving the way for a better understanding of high-performance quantum algorithms for current noisy devices.


\end{document}